\documentclass[a4paper,11pt]{article}
\usepackage{pos}

\usepackage{graphicx}
\usepackage{xcolor}
\usepackage{xspace}
\usepackage{amsmath}

\newcommand{\fbinv} {\mbox{\ensuremath{\,\text{fb}^{-1}}}\xspace}
\newcommand{\GeV}{\ensuremath{\,\mathrm{Ge\hspace{-.08em}V}}\xspace}
\newcommand{\TeV}{\ensuremath{\,\mathrm{Te\hspace{-.08em}V}}\xspace}
\newcommand{\pT}{\ensuremath{p_{\mathrm{T}}}\xspace}
\newcommand{\ttbar}{t\ensuremath{\mathrm{\bar{t}}}\xspace}
\newcommand{\Zprime}{\ensuremath{{\text{Z}\,'}}\xspace}

\renewcommand{\logo}{\relax}

\title{CMS highlights on searches for new physics in final states with jets}

\author*[a]{Emmanouil Vourliotis}
\onbehalf{on behalf of the CMS Collaboration}

\affiliation[a]{University of California, San Diego,\\
                California, United States of America}

\emailAdd{emmanouil.vourliotis@cern.ch}

\abstract{
Many new physics models, e.g., leptoquarks, extra dimensions, extended Higgs sectors, supersymmetric theories, and dark sector extensions, are expected to manifest themselves in the final states with hadronic jets.
Novel experimental techniques, including a dedicated scouting trigger stream and advanced machine learning techniques can be employed to identify such signals.
This talk presents searches in CMS for new phenomena in the final states that include jets, focusing on the most recent results obtained using the full Run-II data-set collected at the LHC.
}

\FullConference{The Eleventh Annual Conference on Large Hadron Collider Physics (LHCP2023)\\
 22-26 May 2023\\
 Belgrade, Serbia\\}

\begin{document}
\maketitle

\section{Introduction} \label{sec:intro}

The Standard Model (SM) of particle physics is the most successful theoretical model to date that explains the interactions between elementary particles with great precision over many orders of magnitude.
Despite its continuing success, there are long-lasting questions that the SM provides no explanation for, e.g., the origin of dark matter or the nonzero mass of the neutrinos.
Motivated by these questions, an abundance of theoretical models have been proposed to answer them and an extensive experimental program has been launched at the CERN Large Hadron Collider (LHC) to unravel new physics phenomena.
Among the different models explored at the LHC, the ones that involve new particles coupling to quarks and/or gluons are of great interest due to their potentially complicated final states with multiple jets.
Two recent searches for new physics in final states with jets performed by the CMS experiment~\cite{CMS1,CMS2} are presented below.
They both utilize the full data set recorded by the CMS experiment during the LHC Run 2 that corresponds to a luminosity of up to $138\fbinv$.

\section{Search for a high-mass dimuon resonance produced in association with b quark jets} \label{sec:ZPrime}


This analysis searches for a new neutral vector boson, \Zprime, with mass $m_\Zprime \geq 350\GeV$~\cite{ZPrime}.
The \Zprime boson is assumed to be produced in association with at least one b quark through its coupling to b and s quarks, and subsequently decays to a pair of muons.
The existence of such a \Zprime boson has implications to low energy $\text{b} \to \text{s} \mu^+ \mu^-$ observables, as explained in~\cite{allanach}.



The analysis selection consists of the requirement for a pair of prompt, isolated muons with transverse momentum $\pT > 53 \GeV$ and pseudorapidity $|\eta| < 2.4$, and the presence of at least one b quark jet ($\pT > 20 \GeV$ and $|\eta| < 2.5$), passing tight b tagging identification criteria, with all other b quark jets required to satisfy relaxed b tagging identification criteria.
The analysis sensitivity is enhanced by the rejection of events in which the minimum invariant mass among any combination of selected muons and b quark jets is less than the mass of the t quark.
This suppresses the \ttbar background by a factor of more than 300, while retaining most of the signal events, especially those with high $m_\Zprime$ hypotheses.
Events are vetoed if the selected muons are back-to-back in 3D space (cosmic muons), if there are additional leptons (electrons or muons) or isolated tracks ($\tau$ leptons) (diboson processes), or if there is significant missing transverse energy in the direction of the selected muons or b quark jets (events with mismeasurements).
The final set of selected events is split into separate categories based on the number of b quark jets, $N_\text{b}$: $N_\text{b} = 1$ and $N_\text{b} \geq 2$.

\begin{figure}[tbh!]
\begin{center}
    \includegraphics[width=0.32\textwidth]{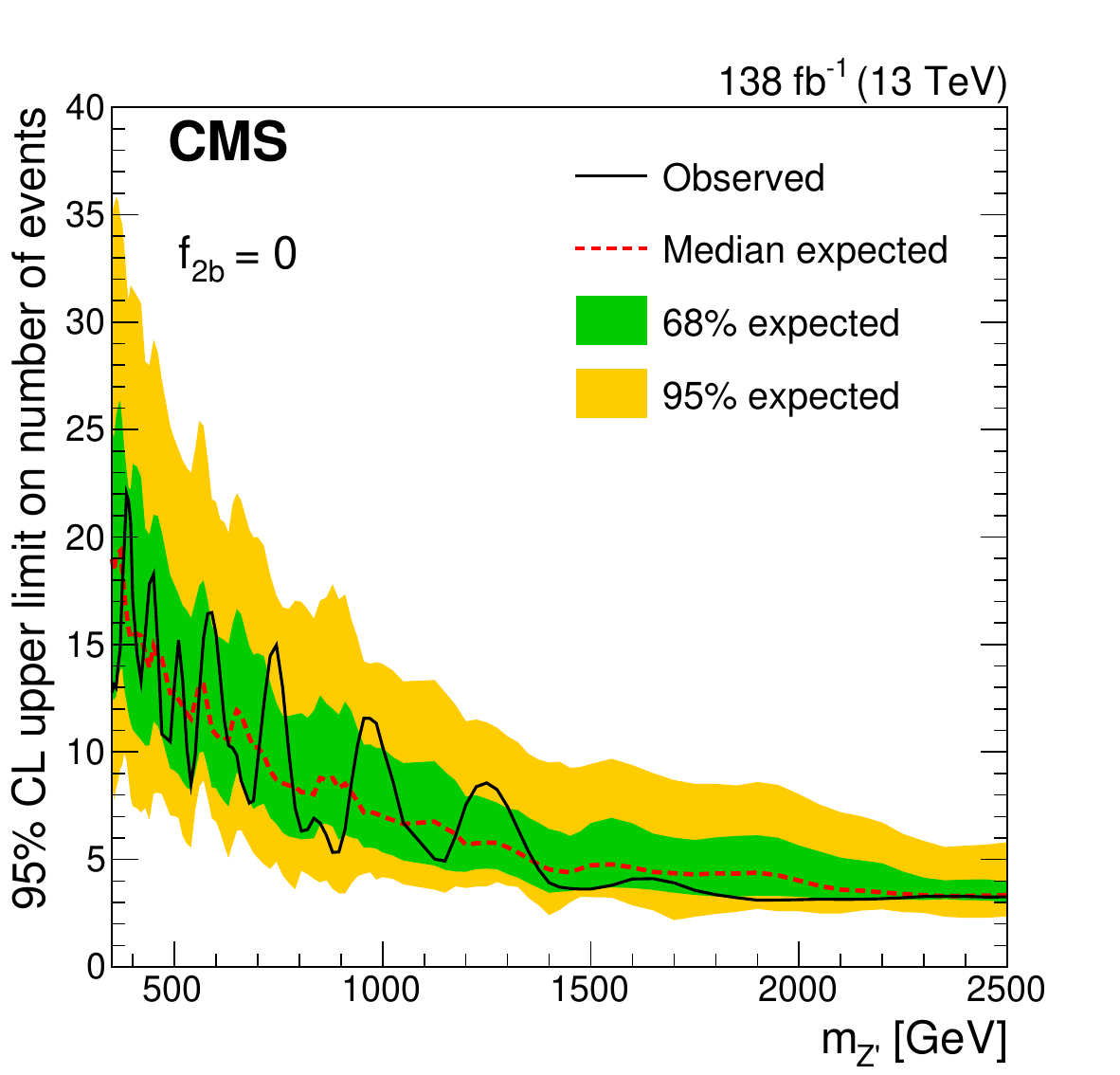}
    \includegraphics[width=0.32\textwidth]{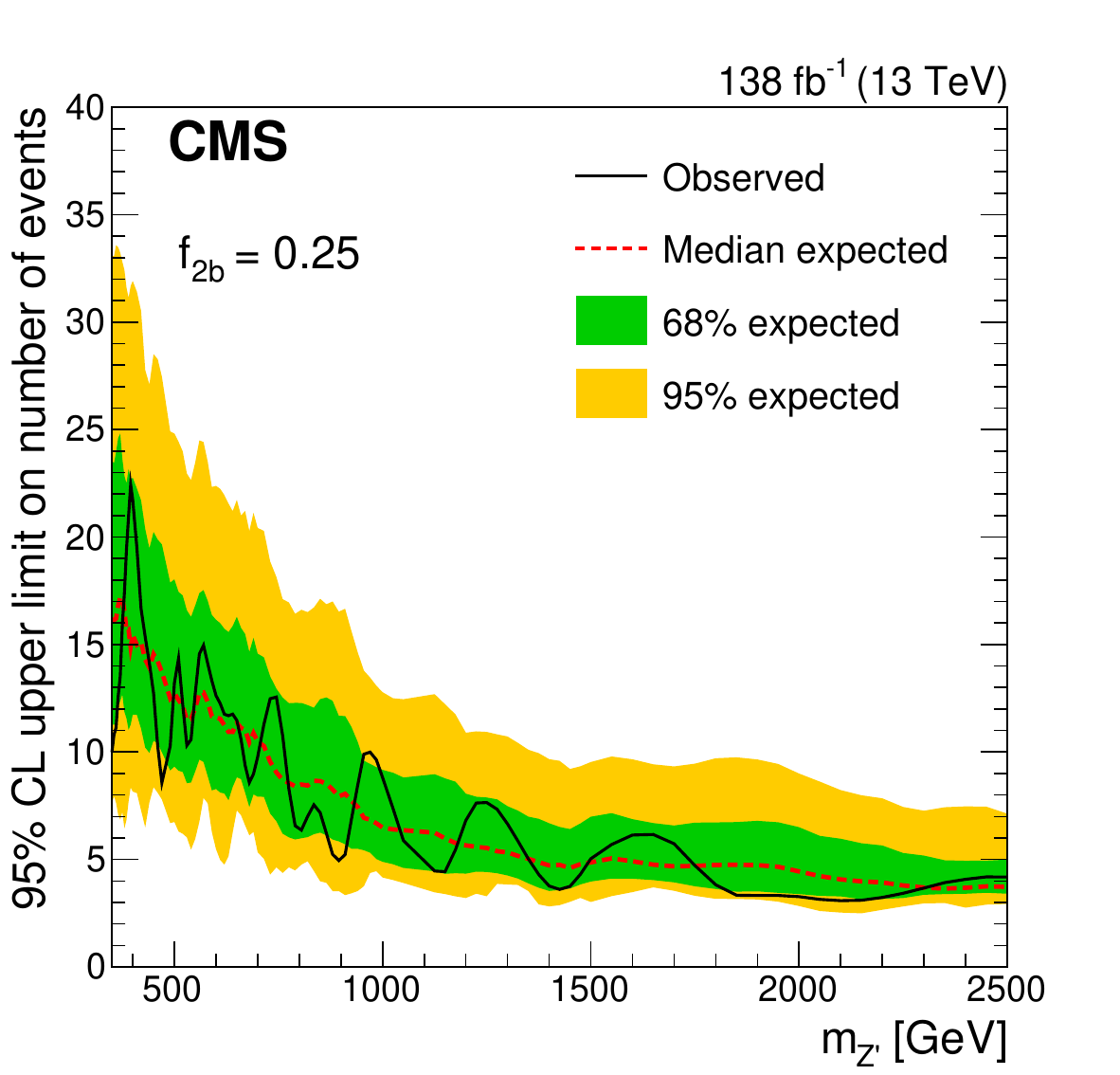}
    \includegraphics[width=0.32\textwidth]{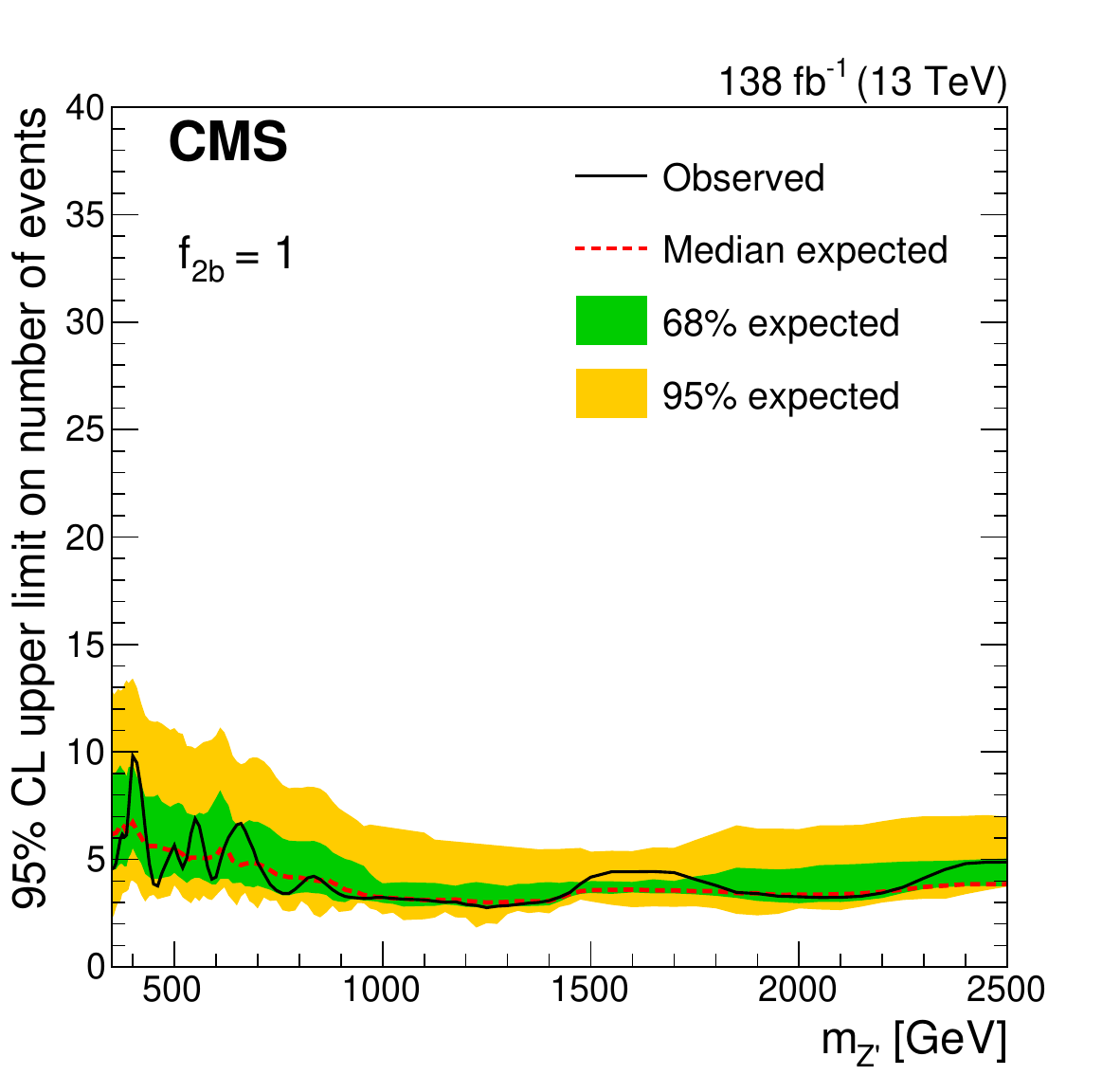}
    \caption{
    Exclusion limits at 95\% CL on the number of BSM events selected by the analysis with $N_\text{b} \geq 1$ as a function of the \Zprime mass, shown for $f_{2\text{b}} =$ 0 (left), 0.25 (middle), and 1 (right).
    The median expected and observed exclusions are represented by the red and black solid lines respectively.
    The green and yellow bands illustrate the 68\% and 95\% uncertainty bands respectively.
    Taken from~\cite{ZPrime}.
    }
    \label{fig:modelIndependentLimits}
\end{center}
\end{figure}
\begin{figure}[tbh!]
\begin{center}
    \includegraphics[width=0.32\textwidth]{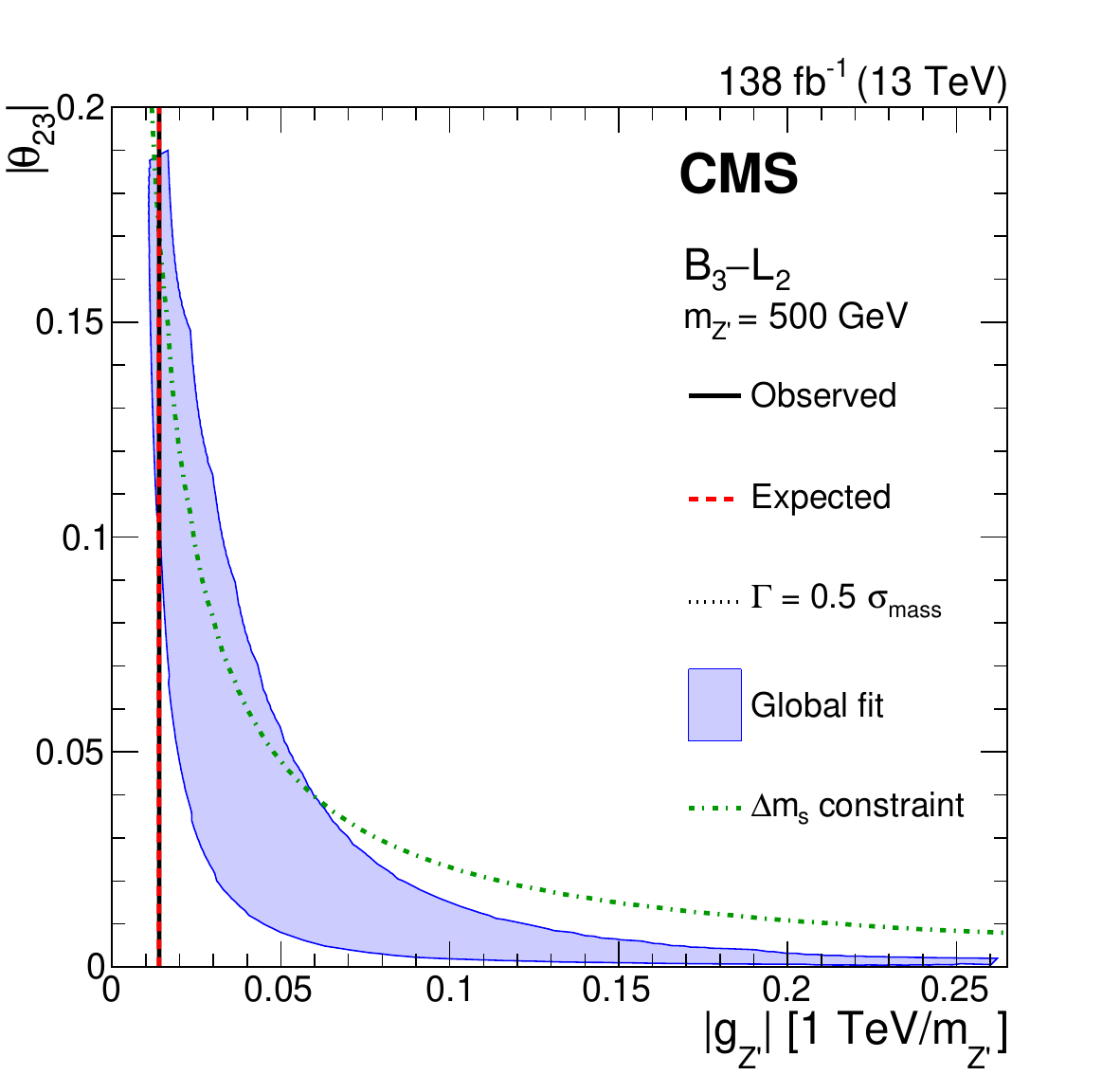}
    \includegraphics[width=0.32\textwidth]{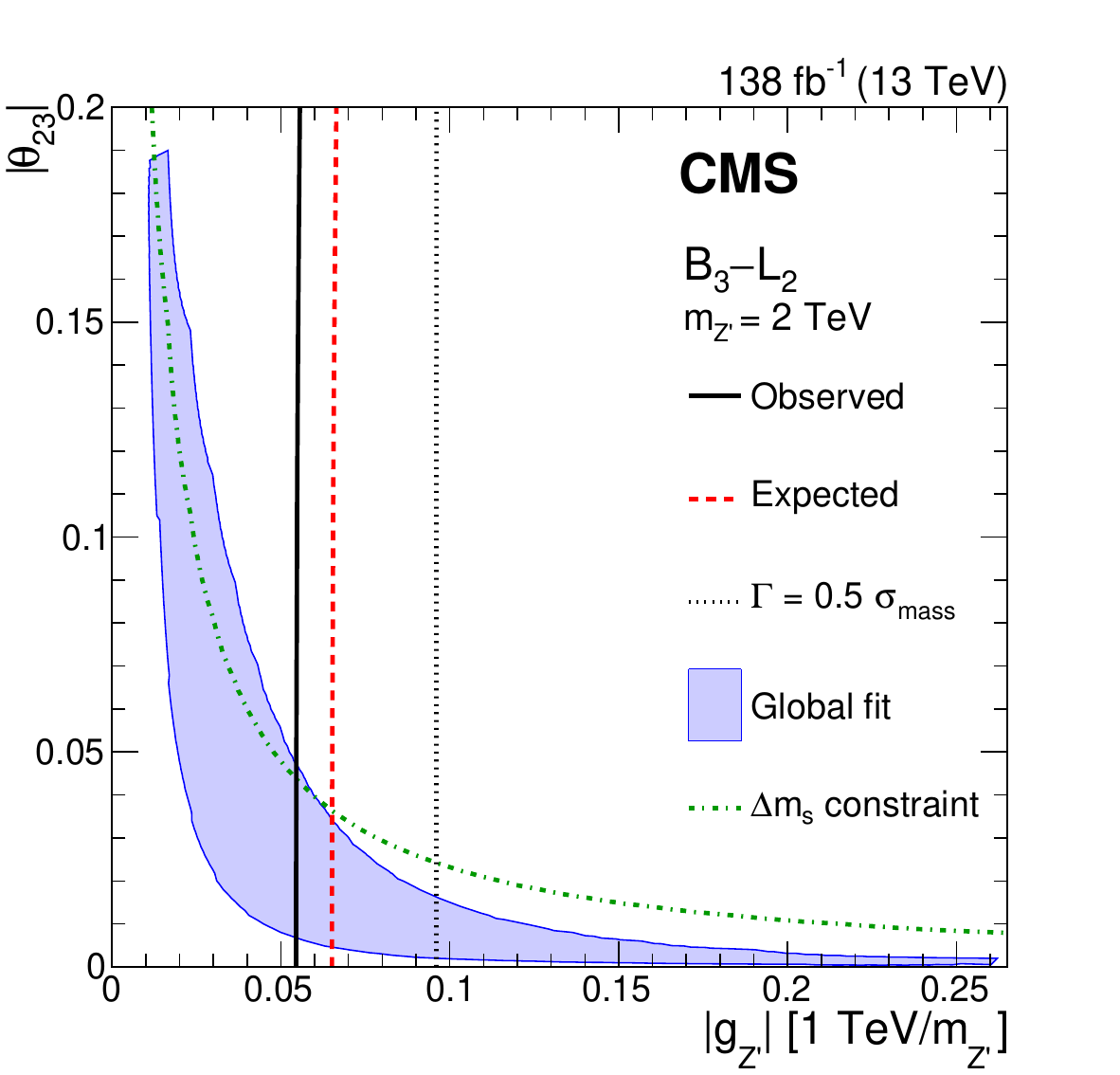}
    \includegraphics[width=0.32\textwidth]{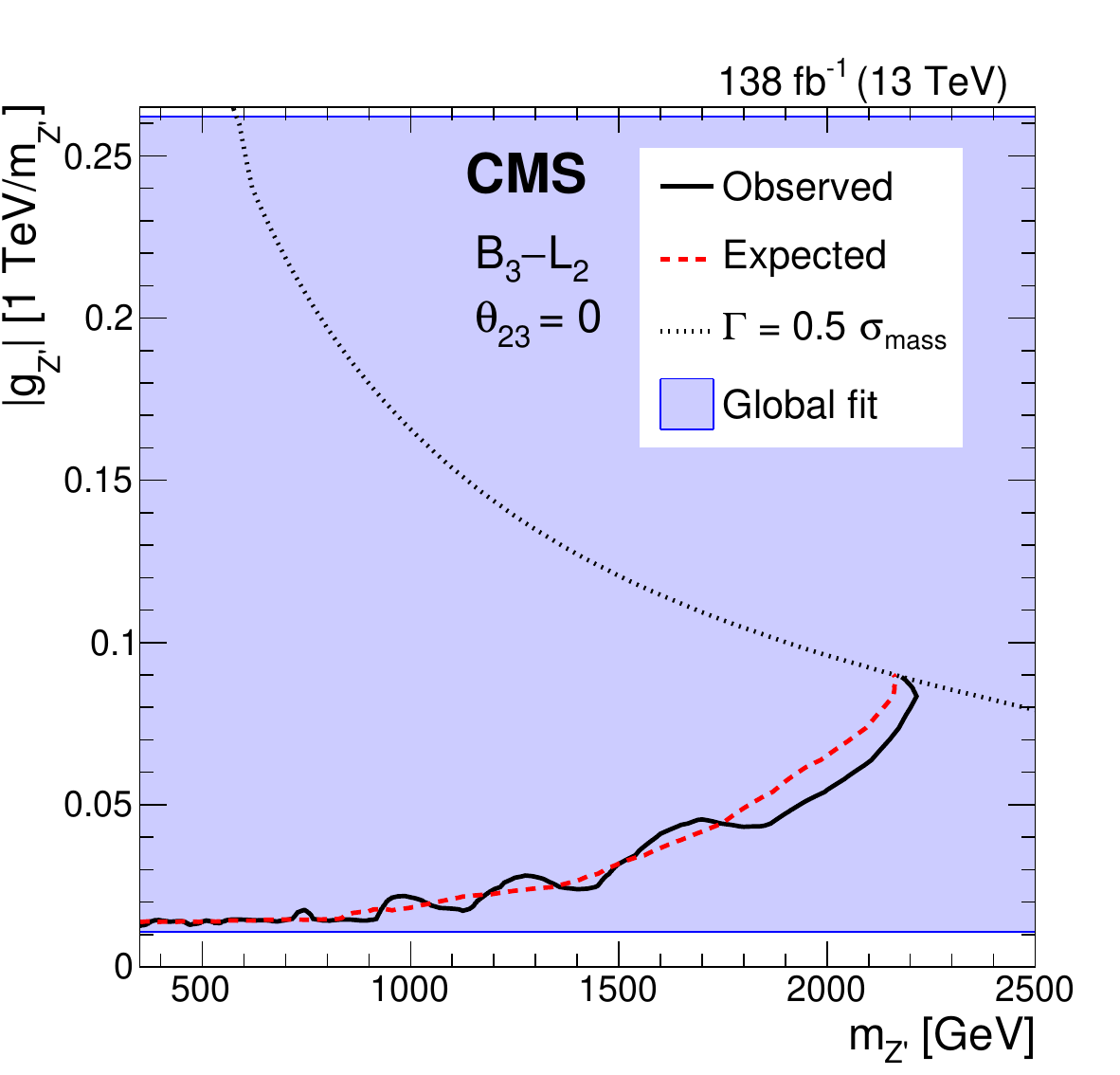}
    \caption{
    Median expected and observed exclusion limits at 95\% CL shown as dashed and solid lines, respectively, for the $\text{B}_3 - \text{L}_2$ model in the $|\theta_{23}| - g_{\Zprime}$ plane, for $m_{\Zprime} = 500\GeV$ (left) and $m_{\Zprime} = 2\TeV$ (middle), and in the $g_{\Zprime} - m_{\Zprime}$ plane, for $|\theta_{23}| = 0$ (right).
    The dotted lines delimit the parameter space in which the \Zprime width is equal to half of the dimuon invariant mass resolution, and, hence, the range of validity for this search.
    The regions enclosed between the solid and the dotted lines are excluded.
    In the cases where lines appear to be missing, they fall outside the range of the plots.
    The shaded blue area illustrates the region preferred by the global fit of Ref.~\cite{allanach} at 95\% CL, and the region above the irregularly dashed line, labeled as ``$\Delta \text{m}_\text{S}$ constraint'', indicates the parameter space incompatible at 95\% CL with the measurement of the mass difference between the neutral $\text{B}_\text{s}$ meson mass eigenstates.
    Taken from~\cite{ZPrime}.
    }
    \label{fig:modelDependentLimits}
\end{center}
\end{figure}

The signal shape is parametrized from simulated samples and the dependence of its parameters is extracted as a function of $m_\Zprime$.
The background is estimated in a fully data-driven way by fitting the dimuon invariant mass distribution in data within windows of width $\pm 10\sigma_{m_\Zprime}$ around the probed $m_\Zprime$.
The uncertainty in the background prediction is estimated using the envelop of the different functions used for the fit (discrete profiling method~\cite{discreteProfiling}).
The results showed no significant excess from the SM.
Given this, model independent limits are set on the number of signal events with $N_\text{b} \geq 2$, as shown in Fig.~\ref{fig:modelIndependentLimits}.
To probe different signal hypotheses, the quantity $f_{2\text{b}}$, defined as the fraction of beyond the SM events at least two b quark jets passing the analysis selection, is varied.
Model dependent limits are also set for the $\text{B}_3 - \text{L}_2$ model of Ref.~\cite{allanach} (Fig.~\ref{fig:modelDependentLimits}).


\section{Search for narrow trijet resonances} \label{sec:trijet}

In models with cascade decays of a heavy resonance, X, decaying to an intermediate resonance Y, and a parton, leading to a 3-parton final state, the ratio $\rho_m = m_\text{Y} / m_\text{X}$ is of central importance, as it defines the experimental signature of the model.
In the case of $\rho_m > 0.8$, the jet from the parton of the $\text{X}\to\text{Y}$ decay is soft, effectively leading to a dijet final state~\cite{inclusiveDijet}.
For $\rho_m < 0.2$, the jets from the Y decay are boosted, resulting in a dijet final state, with one of the jets including two partons, hence having a complicated substructure~\cite{boostedDijet}.
Finally, for intermediate values, $0.2 \leq \rho_m \leq 0.8$, the final state has 3 resolved jets, and can be used to probe compositeness models~\cite{compositeness}. The same final state can be the result of a direct 3-body decay, as predicted in models with extra dimensions~\cite{extraDimensions}.
The first generic search for narrow resonances leading to a trijet final state is described in the following~\cite{trijet}.


The jet selection relies on the reconstruction of exactly three ``wide-jets'':
Three AK4 jets ($\pT > 100\GeV$ and $|\eta| < 2.5$) are considered as ``seeds'' to cluster other nearby jets ($\pT > 30\GeV$ and $|\eta| < 2.5$) within a cone of radius $\Delta R = \sqrt{\Delta\eta^2 + \Delta\phi^2}= 1.1$.
These jets are used to recover energy that may ``leak'' outside of the conventional AK4 jets, which can be a common occurrence for jets originating from high \pT gluons.
Apart from the wide-jet requirement, the selected events are required to satisfy $\min\left(\Delta\eta_\text{jj}\right) < 1.6$ and $\max\left(\Delta\eta_\text{jj}\right) < 3.0$, a selection based on $S/\sqrt{B}$ optimization studies, and $m_\text{jjj} > 1.50~(1.76)\TeV$ for 2016~(2017 and 2018), dictated by the trigger selection.



\begin{figure}[tbh!]
\begin{center}
    \includegraphics[width=0.38\textwidth]{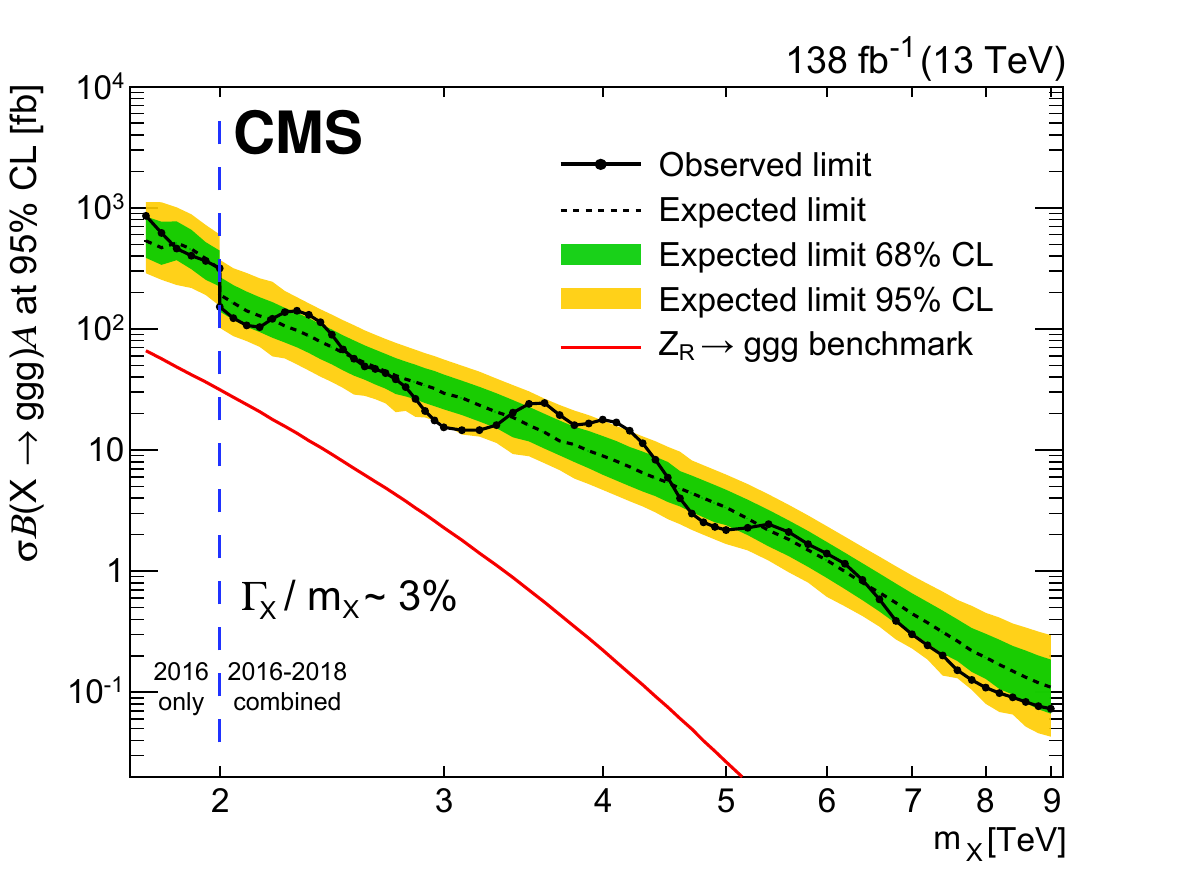}
    \includegraphics[width=0.30\textwidth]{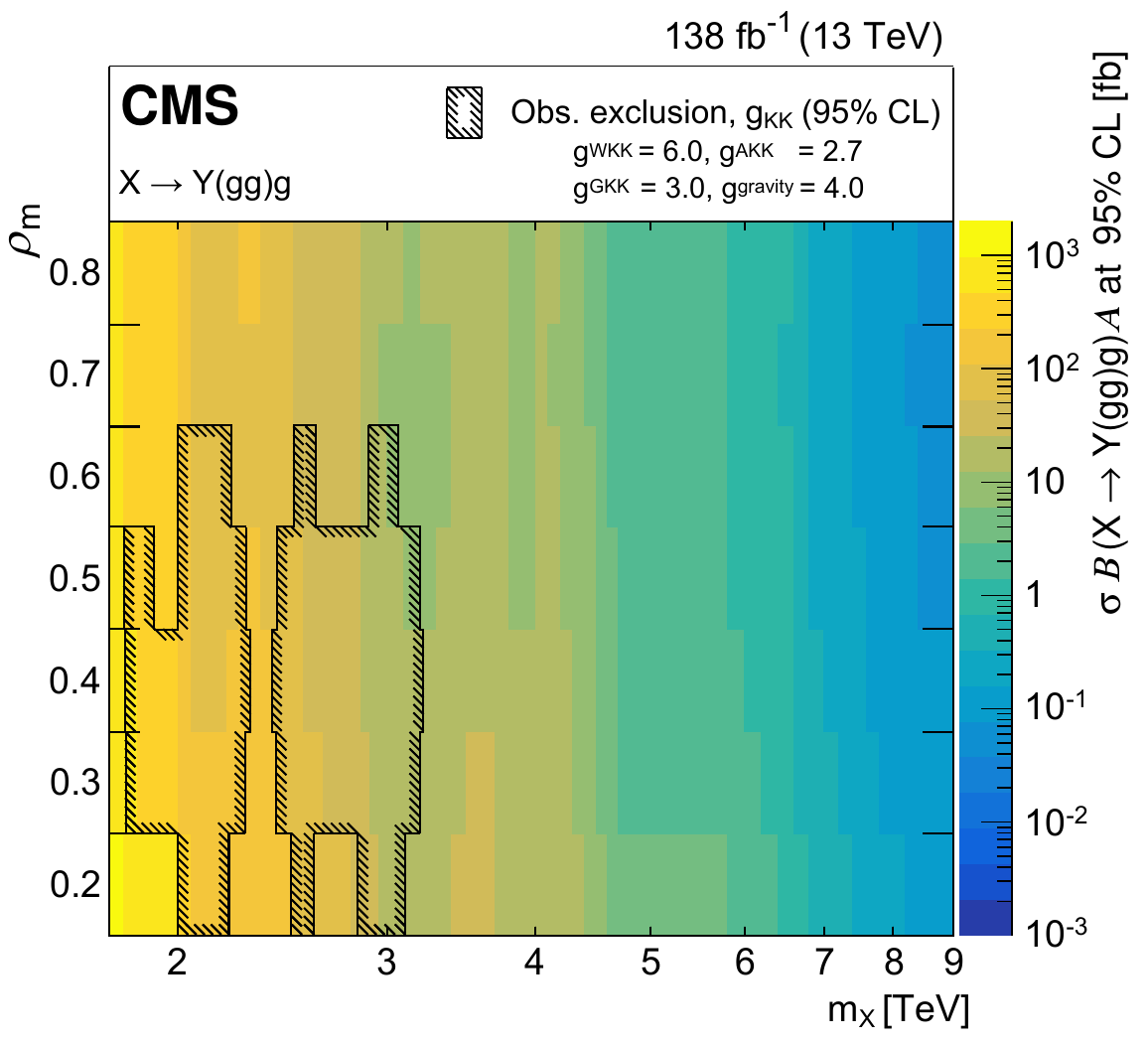}
    \includegraphics[width=0.30\textwidth]{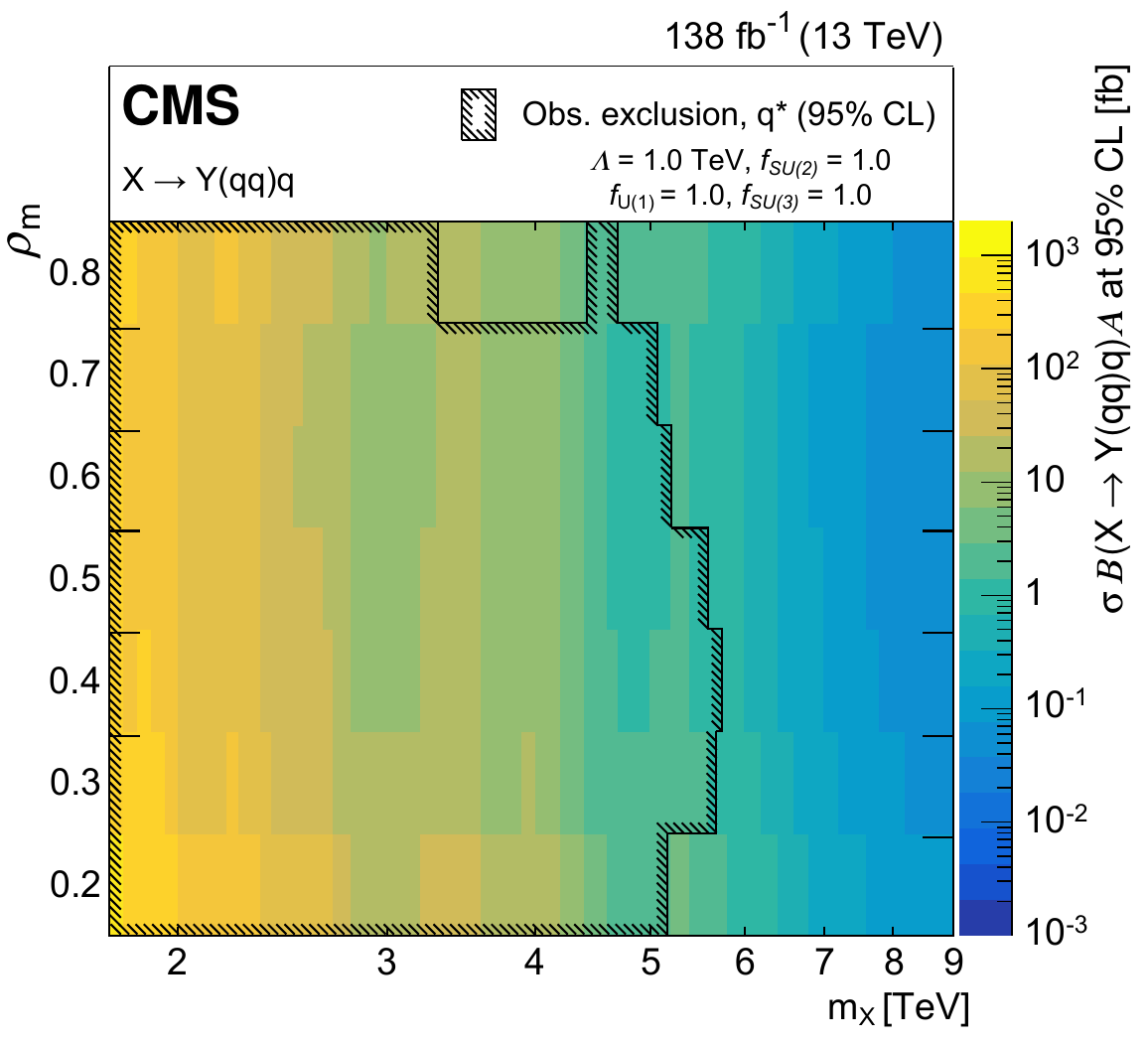}
    \caption{Exclusion limits of the cross section times branching fraction at 95\% CL are shown for the nominal $Z_R$ model (left), the $G_{KK} \rightarrow \phi g \rightarrow (gg)g$ (middle), and $q^* \rightarrow \text{W}\,' q \rightarrow (qq)q$ (right).
    Only 2016 data is used for the derivation of limits below 2\TeV due to the higher trigger thresholds in 2017 and 2018.
    The red curve on the left plot shows the theoretical prediction for a model with SM-like couplings.
    Taken from~\cite{trijet}.}
    \label{fig:trijetLimit}
\end{center}
\end{figure}

The signal extraction is performed using a binned maximum likelihood fit on the trijet mass spectrum, with the binning determined by the $m_\text{jjj}$ resolution, which varies from 2\% to 4\%.
The modeling of the signal is done by fitting the signal shape of simulated samples and interpolating the parameters using the template morphing method. The background estimation relies on constructing the envelope of empirical functions fitted on the inclusive trijet invariant mass distribution in data.
No significant excess was observed, with the maximum local~(global) significance of $2.2~(0.36)\sigma$.
Even though the current dataset has not enough statistical power to reach exclusion for the $Z_R$ model described in Ref.~\cite{extraDimensions} (Fig.~\ref{fig:trijetLimit}, left), exclusion limits up to 3.1 and $6.0\TeV$ are set for the $G_{KK} \to \phi g \to ggg$ and $q^* \to \text{W}\,' q \to qqq$ models, summarized in Ref.~\cite{compositeness}, respectively.

\newpage

\bibliographystyle{JHEP}
\bibliography{LHCP2023}

\providecommand{\href}[2]{#2}\begingroup\raggedright\begin{thebibliography}{10}

\bibitem{CMS1}
{CMS Collaboration}, \emph{{The CMS experiment at the CERN LHC}}, \href{https://doi.org/10.1088/1748-0221/3/08/S08004}{\emph{{Journal of Instrumentation}} {\bfseries 3} (2008) S08004}.

\bibitem{CMS2}
{CMS Collaboration}, \emph{{Development of the CMS detector for the CERN LHC Run 3}},  2023.
\newblock \href{https://arxiv.org/abs/2309.05466}{{\ttfamily 2309.05466}}, {Accepted by Journal of Instrumentation}.

\bibitem{ZPrime}
{CMS Collaboration}, \emph{{Search for a high-mass dimuon resonance produced in association with b quark jets at $\sqrt{s} = 13\TeV$}}, \href{https://doi.org/10.1007/jhep10(2023)043}{\emph{{Journal of High Energy Physics}} {\bfseries 2023} (2023) 043}.

\bibitem{allanach}
B.~Allanach and J.~Davighi, \emph{{The Rumble in the Meson: a leptoquark versus a \Zprime to fit $\text{b} \to \text{s} \mu^+ \mu^-$ anomalies including 2022 LHCb $\textrm{R}_{K^{(\ast)}}$ measurements}}, \href{https://doi.org/10.1007/JHEP04(2023)033}{\emph{{Journal of High Energy Physics}} {\bfseries 04} (2023) 033}.

\bibitem{discreteProfiling}
P.~Dauncey, M.~Kenzie, N.~Wardle and G.~Davies, \emph{Handling uncertainties in background shapes: the discrete profiling method}, \href{https://doi.org/10.1088/1748-0221/10/04/P04015}{\emph{Journal of Instrumentation} {\bfseries 10} (2015) P04015}.

\bibitem{inclusiveDijet}
{CMS Collaboration}, \emph{{Search for high mass dijet resonances with a new background prediction method in proton-proton collisions at $\sqrt{s} = 13\TeV$}}, \href{https://doi.org/10.1007/JHEP05(2020)033}{\emph{{Journal of High Energy Physics}} {\bfseries 2020} (2020) 33}.

\bibitem{boostedDijet}
{CMS Collaboration}, \emph{{Search for high-mass resonances decaying to a jet and a Lorentz-boosted resonance in proton-proton collisions at $\sqrt{s} = 13\TeV$}}, \href{https://doi.org/10.1016/j.physletb.2022.137263}{\emph{{Physics Letters B}} {\bfseries 832} (2022) 137263}.

\bibitem{compositeness}
{K. Huitu and J. Maalampi and A. Pietilä and M. Raidal}, \emph{{Doubly charged Higgs at {LHC}}}, \href{https://doi.org/10.1016/s0550-3213(97)87466-4}{\emph{{Nuclear Physics B}} {\bfseries 487} (1997) 27}.

\bibitem{extraDimensions}
{Kaustubh S. Agashe and Jack H. Collins and Peizhi Du and Sungwoo Hong and Doojin Kim and Rashmish K. Mishra}, \emph{{{LHC} signals from cascade decays of warped vector resonances}}, \href{https://doi.org/10.1007/jhep05(2017)078}{\emph{{Journal of High Energy Physics}} {\bfseries 2017} (2017) }.

\bibitem{trijet}
{CMS Collaboration}, \emph{{Search for narrow trijet resonances in proton-proton collisions at $\sqrt{s} = 13\TeV$}},  2023.
\newblock \href{https://arxiv.org/abs/2310.14023}{{\ttfamily 2310.14023}}, {Submitted to Physical Review Letters}.

\end{thebibliography}\endgroup

\end{document}